\documentclass[12pt,preprint]{aastex}
\usepackage{color}
\usepackage{graphicx}
\usepackage{natbib}
\usepackage{epsfig}

\begin{document}

\title{TeV Gamma-Ray Sources from a Survey of the Galactic Plane with Milagro}

\author{
A.~A.~Abdo,\altaffilmark{1}
B.~Allen,\altaffilmark{2}
D.~Berley,\altaffilmark{3}
S.~Casanova,\altaffilmark{4}
C.~Chen,\altaffilmark{2}
D.~G.~Coyne,\altaffilmark{5}
B.~L.~Dingus,\altaffilmark{4}
R.~W.~Ellsworth,\altaffilmark{6}
L.~Fleysher,\altaffilmark{7}
R.~Fleysher,\altaffilmark{7}
M.~M.~Gonzalez,\altaffilmark{8}
J.~A.~Goodman,\altaffilmark{3}
E.~Hays,\altaffilmark{3}
C.~M.~Hoffman,\altaffilmark{4}
B.~Hopper,\altaffilmark{3}
P.~H.~H\"untemeyer,\altaffilmark{4}
B.~E.~Kolterman,\altaffilmark{7}
C.~P.~Lansdell,\altaffilmark{3}
J.~T.~Linnemann,\altaffilmark{1}
J.~E.~McEnery,\altaffilmark{9}
A.~I.~Mincer,\altaffilmark{7}
P.~Nemethy,\altaffilmark{7}
D.~Noyes,\altaffilmark{3}
J.~M.~Ryan,\altaffilmark{10}
P.~M.~Saz~Parkinson,\altaffilmark{5}
A.~Shoup,\altaffilmark{11}
G.~Sinnis,\altaffilmark{4}
A.~J.~Smith,\altaffilmark{3}
G.~W.~Sullivan,\altaffilmark{3}
V.~Vasileiou,\altaffilmark{3}
G.~P.~Walker,\altaffilmark{4}
D.~A.~Williams,\altaffilmark{5}
X.~W.~Xu\altaffilmark{4}
and
G.~B.~Yodh\altaffilmark{2}}

\altaffiltext{1}{ Michigan State University, East Lansing, MI}
\altaffiltext{2}{ University of California, Irvine, CA}
\altaffiltext{3}{ University of Maryland, College Park, MD}
\altaffiltext{4}{ Los Alamos National Laboratory, Los Alamos, NM}
\altaffiltext{5}{ University of California, Santa Cruz, CA}
\altaffiltext{6}{ George Mason University, Fairfax, VA}
\altaffiltext{7}{ New York University, New York, NY}
\altaffiltext{8}{ Instituto de Astronomia, Universidad Nacional Autonoma de Mexico, D.F., MEXICO}
\altaffiltext{9}{ NASA Goddard Space Flight Center, Greenbelt, MD}
\altaffiltext{10}{ University of New Hampshire, Durham, NH}
\altaffiltext{11}{ Ohio State University, Lima, OH}

\begin{abstract}
A survey of Galactic gamma-ray sources at a median energy of $\sim$20 TeV has been performed 
using the Milagro Gamma Ray Observatory.
Eight candidate sources of TeV emission are detected with pre-trials significance
$>4.5\sigma$  in the region of Galactic longitude
$l\in[30^\circ,220^\circ]$ and latitude $b\in[-10^\circ,10^\circ]$. 
Four of these sources, including the Crab nebula and the recently published MGRO J2019+37, 
are observed with significances $>4\sigma$ after accounting for the trials involved in searching the 
3800 square degree region.
All four of these sources are also coincident with EGRET sources.  Two of the lower significance sources are 
coincident with EGRET sources and one of these sources is Geminga.  The other two candidates are
in the Cygnus region of the Galaxy.
Several of the sources appear to be spatially extended. The fluxes of the sources at 20 TeV
range from $\sim25\%$ of the Crab flux to nearly as bright as the Crab.
\end{abstract}


\section{Introduction}

A survey of the entire Northern Hemisphere sky for sources of TeV gamma rays 
has been performed using the Milagro Gamma Ray Observatory \citep{atkins_allsky}.
This paper reports on observations of sources in the region of Galactic longitude
$l\in[30^\circ,220^\circ]$ and latitude $b\in[-10^\circ,10^\circ]$. This survey is
at a higher energy of $\sim20$ TeV and in a different region than the survey performed by
the HESS atmospheric Cherenkov telescope (ACT) above 200 GeV.  The HESS survey covered 
$l\in[-30^\circ,30^\circ]$ and $b\in[-2^\circ,2^\circ]$ and resulted in the detection
of 14 new sources \citep{2006ApJ...636..777A}. At even lower energies, 
EGRET on the Compton Gamma Ray Observatory detected 80 sources
above 100 MeV \citep{3EG} and 28 sources above 1 GeV \citep{GEV} within 10 deg of the
Galactic plane. Fourteen of these GeV sources are in the region surveyed by Milagro.
Most of the EGRET sources and several of the TeV sources are unidentified
without counterparts at lower energies. However, the improved localizations of HESS have
led to identification of  supernova remnants
(SNR)\citep{2004Natur.432...75A}, pulsar wind nebulae (PWN) \citep{2005A&A...442L..25A},
and molecular clouds \citep{2006Natur.439..695A}
as emitters of TeV gamma rays.

In the Milagro data, four sources (including the Crab) are detected with post-trials significance 
greater than 4$\sigma$, and four additional lower significance candidates are identified. 
Six of these eight TeV excesses are coincident with the locations of EGRET sources. Many of 
these sources have stringent upper limits at TeV energies, such as those from the 
Whipple observatory \citep{2005ApJ...624..638F}. However, these upper limits weaken for
extended sources by the ratio of the radius of the 
source to the  angular resolution. The angular resolution
of Milagro is approximately an order of magnitude larger
than ACTs, but is less than half that of EGRET, resulting in little reduction in 
sensitivity for sources up to $\sim2^\circ$ diameter.
The TeV flux, spatial morphology, and potential counterparts 
of these new sources are discussed below.

\section{Analysis}

The analysis was performed on 2358 days of data collected by Milagro during the
operational period from July 19, 2000 through Jan 1, 2007. Data from the last 3 years 
of operation were collected after the completion of the outrigger array, which 
substantially increased the sensitivity of Milagro, particularly at high energies. 
The Milagro data were analyzed using the method described in \citet{2007ApJ...658L..33A}, in which
the events are weighted based on the gamma/hadron separation parameter ($A_{4}$).
The signal and background maps are smoothed with the point spread function (PSF),
which varies based on the number of hits in the events. The statistical significance 
of the excess or deficit is computed using eqn. 17 in \citet{LiMa}.

The Milagro detector is located at latitude $36^\circ$ N, where the Galactic center and
the central Galactic bulge are outside the field of view.  In this survey of the Northern Galactic
plane, only events with zenith angle less than $45^\circ$ are included, which 
covers declinations north of $\delta=-7^\circ$. In Galactic coordinates, this 
region is longitude $l\in[30^\circ,220^\circ]$ and latitude $b\in[-10^\circ,10^\circ]$,
which subtends 9.2\% of the 
entire sky and covers more than half of the Galactic plane. The energy threshold and 
sensitivity of Milagro vary with zenith angle. Table 1 shows the median energy and 
relative sensitivity of the instrument as a function of declination for an 
assumed differential photon power law spectrum of spectral index $\alpha=-2.3$. 

The background at a given location in celestial coordinates is found use the method
described in \citet{atkins_crab} with the modification that the events are weighted. 
The signal map is searched for excesses over the background map with a statistical
significance $>4.5\sigma$.
When a source candidate is found, the region in the vicinity of the candidate is 
removed to prevent double counting, and the search is repeated. 
With this automated procedure, two nearby sources may be identified as a
single source candidate or a single extended source may be broken up into
multiple candidates. 

The event excess in the vicinity of each source candidate is fit to a two-dimensional 
Gaussian to determine the location, spatial extent and flux. The average angular 
resolution\footnote{To be consistent with HESS and EGRET, the angular resolution
is described by the radius that contains 68\% of the signal events from a
point source. In previous publications, the sigma of the Gaussian function was
referenced, which is $\sim0.7^\circ$.}
of these weighted events is $1.1^\circ$ and has been measured with 
observations of the Crab.  Astrophysical sources are better described by a top hat
function than a Gaussian; however, a Gaussian 
approximates a top hat of less than a few degrees diameter that is convolved with
the Milagro PSF.  Therefore, the Gaussian sigma is used to constrain the diameter 
of the top hat function that best fits the angular extent of the source.
 
The excess above the assumed isotropic background is calculated from the
volume of the fit two-dimensional Gaussian.  This calculation results in a
larger uncertainty than implied by the significance of the detection, but
properly accounts for the flux of extended sources. 
The excess is converted to a flux with a Monte Carlo simulation of 
extensive air showers using CORSIKA \citep{corsika}
and of the Milagro detector using GEANT4 \citep{geant}. The source fluxes are 
computed assuming a differential photon spectrum of a power law with spectral index 
$\alpha=-2.3$ with no cutoff. This is the average spectrum for Galactic sources 
observed by HESS \citep{2006ApJ...636..777A}. The flux is quoted at
20 TeV, which is approximately the median energy of the gamma rays 
from the sources detected by Milagro, as seen in table 1.
The median energy of Milagro varies with source spectra and declination, however 
the uncertainty in the flux at 20 TeV is relatively insensitive to the
assumed spectrum and varies by only $<20\%$ when $\alpha$ varies from 
$-2.0$ to $-2.6$.

\section{Results}

\begin{figure*}[width=4.5in,p]

%\plotone{galplane_3panel_blue.eps}
\plotone{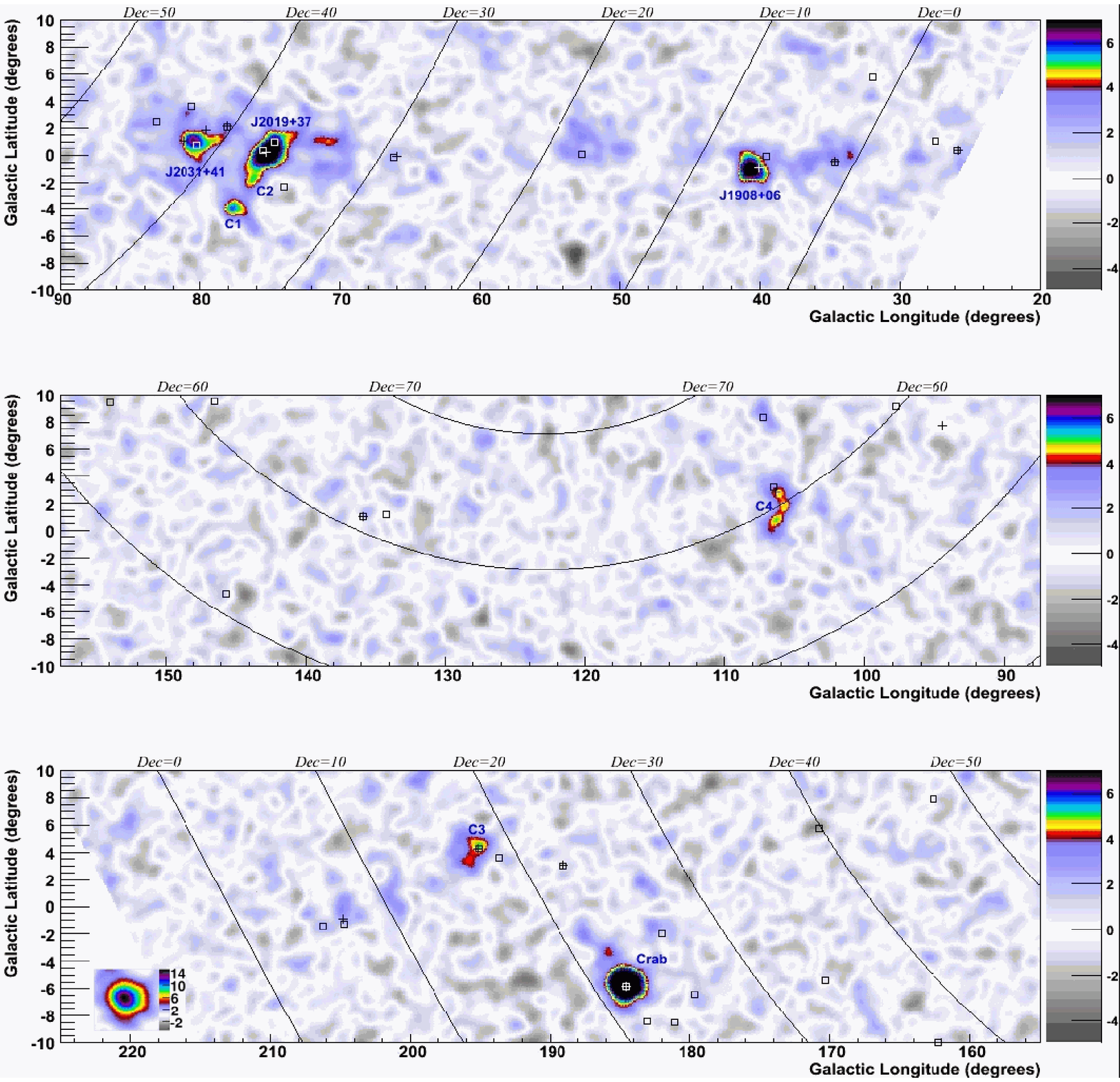}
\caption{Significance map of the Galactic plane. The color code shows the pre-trials 
significance in this PSF-smoothed map. The maximum positive value of the color code saturates at 7$\sigma$
although three of the gamma-ray sources are detected with much higher statistical significance. The Crab 
image is inset with the same x and y scale in the bottom left as an indication of the PSF. 
Boxes (crosses) indicate the locations of the EGRET 3EG (GEV) sources. }
\label{fig:galactic_plane}
\end{figure*}

Figure \ref{fig:galactic_plane} shows a PSF-smoothed map of the Galactic plane, with the color scale
indicating the statistical significance of the Milagro excess or deficit at each point.  
Table \ref{tab:galactic_sources} gives the locations,
fluxes, statistical significances, angular extents, and counterparts for the 
eight source candidates identified with a pre-trials significance in the PSF-smoothed map
of $>4.5\sigma$.  
Because several of the candidates are extended beyond the PSF of Milagro, the maximum pre-trials 
significance for a search with a larger bin of size
$3^\circ \times 3^\circ$ is also given in Table \ref{tab:galactic_sources}.
However, the post-trials significances are not based on this {\em a posteriori} observation, but are calculated from the 
PSF-smoothed pre-trials significances. A Monte Carlo 
simulation is used to account for the trials involved in 
searching this 3800 square degree region. This simulation predicts that 4\% of such searches would
result in at least one source with 
$> 4.5 \sigma$ pre-trials significance anywhere in the region due to background fluctuation.
The list of eight candidates includes the Crab and MRGO 2019+37, which was previously reported 
\citep{2007ApJ...658L..33A}.
Excluding these previously known gamma-ray sources, six new candidates are identified.
The two most significant of these, MGRO J1908+06 and MGRO J2031+41, 
exceed $4.5\sigma$ after accounting for trials. Therefore, the four most significant sources in Table 1
are considered definitive TeV gamma-ray source detections. 
The remaining four source candidates, labeled as C1-C4, have post-trials significances less 
than $4.5\sigma$ and are regarded as lower confidence detections. 

The non-isotropic Galactic diffuse gamma-ray background will contribute to the observed flux of a source.
Diffuse emission is particularly bright in the Cygnus region, where five of the eight candidate sources
are located. The GALPROP model \citep{galprop} predicts that the TeV gamma-ray diffuse
emission flux is 3 to 7 times lower than the observed flux in the large Cygnus
region defined as $l\in[65^\circ,85^\circ]$ and $b\in[-2^\circ,2^\circ]$\citep{2007ApJ...658L..33A}.  
For individual sources near high concentrations of matter, the diffuse
emission will be a larger fraction of the quoted flux.  However, due to uncertainties in the model
predictions, the diffuse emission was not subtracted from the source fluxes.

The {\bf Crab Nebula}, a standard reference source for TeV astronomy, is detected at 15.0$\sigma$.
The flux derived from the Milagro data is consistent, within errors, 
with the flux results obtained from atmospheric Cherenkov telescopes \citep{2004ApJ...614..897A}. 
The fit location is $0.11^\circ$ from the pulsar location, which is consistent with the statistical 
error.

{\bf MGRO J2019+37} is one of five Milagro sources in the Cygnus region of the Galactic plane.
It is the most significant source detection by Milagro after the Crab and was discussed in depth in
\citet{2007ApJ...658L..33A}. More data have been analyzed since the publication of that paper, 
so the flux and significance in Table \ref{tab:galactic_sources} have been updated.

{\bf MGRO J1908+06} is observed with a pre-trials significance of 8.3$\sigma$, with a flux that is
$\sim$80\% of the Crab flux. This location is the closest of the eight candidates to the inner Galaxy, where the 
diffuse emission is expected to increase. This source is coincident with GEV J1907+0557 and with the
bright radio, shell-type, SNR G40.5-0.5\citep{Green}. GEV J1097+0557 was observed for 
87 hours by the HEGRA TeV Observatory, resulting in
an upper limit for a point source of 2.6\% of the Crab flux at 700 GeV \citep{2005A&A...439..635A}.
The Milagro data are consistent both with a point source and with an extended source with diameter 
up to $<2.6^\circ$. The Milagro data are consistent both with a point source and with an extended 
source of diameter up 2.6$^\circ$. The Milagro flux and HEGRA upper limit taken together imply a
harder spectrum than other Galactic sources detected in this energy range by HESS.  However, if 
the source is extended, the HEGRA upper limit would be increased, allowing for a spectrum typical
of other TeV sources.
The Tibet Air Shower Observatory has a similar energy threshold and angular resolution as Milagro, 
and reports a location within $0.9^\circ$ of {\bf MGRO J1908+06} as one of eight locations in their survey 
of the Northern Hemisphere sky that is above 4.5$\sigma$\citep{tibet}.

{\bf MGRO J2031+41} is observed with a pre-trials significance of 6.6$\sigma$ and 
is located in the area with the largest concentration of molecular and atomic gas in the
Cygnus region. Its location is consistent with EGRET sources GEV J2035+4214 and 3EG J2033+4118 
and with TEV J2032+413 \citep{2005A&A...431..197A}. The flux of TeV J2032+413 is measured up to $\sim$10 TeV,
and is only about one third of the Milagro flux when extrapolated to 20 TeV. The low energy counterpart 
for TEV J2032+413 is unclear, but several possibilities have recently been postulated 
\citep{CygTeVCounter1,2007PhRvD..75f3001A}.
The spatial extent of the Milagro detection, at $3.0^\circ \pm 0.9^\circ$, is much larger than the 
few arcminute extent of TEV J2032+413. There must be
another source or sources contributing to the Milagro excess.

There are two source candidates in the Cygnus region -- {\bf C1} and {\bf C2}
-- that are less significant and have no obvious EGRET, PWN, or SNR counterparts.
{\bf C1} is located farthest from the Galactic plane at b = $-3.9^\circ$.
{\bf C2} may be an extension of {\bf MGRO J2019+37} but is $2.2^\circ$ away.
{\bf C2} is not well fit by a two dimensional Gaussian.
For this source, the bin in the PSF-smoothed map with the highest excess is used to determine
the best location, and the flux is obtained from the excess in a $2.0^\circ \times 2.0^\circ$ angular bin.

{\bf C3} is positionally consistent with Geminga, which is the brightest EGRET source in the
Northern Hemisphere sky. 
Geminga is a pulsar that is only 160 pc away and is $\sim$300,000 years old \citep{1996ApJ...461L..91C}.
The 5.1$\sigma$ source detected by Milagro is consistent with the
pulsar location and is extended with a diameter of $2.8^\circ \pm 0.8^\circ$.
The significance increases to $5.9\sigma$ in a $3^\circ\times3^\circ$ bin.
The significance of the Milagro excess at the location of the pulsar is 4.9$\sigma$ in the PSF-smoothed map.
Only pulsed emission was detected by EGRET, but a PWN has been observed in X-rays \citep{2003Sci...301.1345C} that
delineates the bow shock created by the pulsar's motion.
The diameter of the excess implies an $\sim8$ pc source extent, which is consistent with the
observations by HESS of more distant PWN \citep{2005A&A...442L..25A}. Upper limits on the TeV flux
from Geminga of $\sim100$ mCrab \citep{1999A&A...346..913A, 1993A&A...274L..17A, vish} assumed the emission
was from a point source and were set at a much lower energy than this observation.
If the spectral index is $\sim-2.3$ or harder, or if the source is extended,
the Milagro detection is consistent with the previously reported limits.

{\bf C4} is the least significant source in table \ref{tab:galactic_sources} at 5.0$\sigma$.
However, as seen in Figure 1, the source appears very elongated and the significance increases to
$6.3\sigma$ with the larger $3^\circ \times 3^\circ$
bin. The source location is consistent with 3EG J2227+6122, GEV J2227+6106, and the SNR G106.6+2.9
with the accompanying Boomerang PWN.  This PWN is predicted by \citet{2005JPhG...31.1465B}
to be the third brightest TeV PWN in the Northern sky, surpassed only by the Crab and PWN G75.2+0.1, which
both have high significance Milagro counterparts.

\section{Discussion}

A survey of $190^\circ$ of the Galactic plane at $\sim20$ TeV has been performed with Milagro 
with a sensitivity from 3 to 6 $\times 10^{-15}$ (TeV$^{-1}$cm$^{-2}$s$^{-1})$.
Eight sources are identified with $>4.5\sigma$ pre-trials significance above the 
isotropic cosmic ray background. All four of the high confidence sources and two of the lower significance 
candidates are associated with EGRET GeV sources, of which there are only 14 in the search region. 
Four of these six sources appear extended, and the significance of the other two lower significance sources increases
substantially with a larger bin of size $3^\circ\times3^\circ$. 
The significance for a true point source would be expected to decrease slowly for larger bin sizes 
due to the addition of gamma rays beyond the edge of the optimal bin. On the other hand, the 
significance of a statistical fluctuation of background would be expected to decrease inversely 
proportional to the bin radius.
The remaining two lower confidence source candidates are within the Cygnus region, 
where the diffuse emission is high and there is a large concentration of EGRET sources.  
The differential photon spectrum that connects 
the fluxes observed by EGRET at 1 GeV with those observed at 20 TeV 
by Milagro is described by a power law with spectral index $\alpha\sim-2.3$, except for
Geminga which is steeper. 
This spectrum implies a break from the harder spectra observed for most of these
sources by EGRET above 100 MeV. 
Four of these sources are associated with PWNe, of which only 11 
are known in this region \citep{PWN_Roberts}, and one is associated with a SNR. Other
members of these classes of sources have also been detected by HESS. 
The angular extents of the Milagro sources are large, as
would be expected if the sources are nearby, like Geminga. The enhanced sensitivity of GLAST
should allow these objects to be resolved at energies above a few GeV.  In addition, a Milagro
energy spectrum analysis is being developed that will measure spectra of these sources up to 50 TeV. 
The morphology at different energies provides important clues about the energy of the particles 
producing the radiation and the energy loss mechanisms of these particles. 
Combining spatial resolution over a broad range of energies is key to understanding the
Galactic sources that emit the highest energy gamma rays.

\begin{acknowledgments}

We thank Scott Delay, Michael Schneider, and Owen Marshall for their dedicated efforts
on the Milagro experiment. We also gratefully acknowledge the financial support of
the National Science Foundation (under grants 
PHY-0245234, %UCI; previous is PHY-0070933
-0302000, %UMD
-0400424, %Milagro Operations; previous is PHY-0075326
-0504201, %NYU current; previous are PHY-9901496 & PHY-0206656
-0601080, %UCSC; previous are PHY-0070927 & PHY-0245143
and
ATM-0002744),  %UNH
the Department of Energy (Office of High Energy Physics), Los
Alamos National Laboratory, the University of California,
and the Institute for Geophysics and Planetary
Physics at Los Alamos National Laboratory.

\end{acknowledgments}

\newpage

\begin{deluxetable}{cccc}
\tablewidth{2.5in}
\tabletypesize{\scriptsize}  % can be \small \tiny \scriptsize and \footnotesize 
\tablecaption{Sensitivity and Energy Response for a $\frac{dN}{dE}\propto E^{-2.3}$ spectrum\label{tab:energy_and_sensitivity}}
\tablehead{
\colhead{\bf Dec.} &
\colhead{\bf Flux\tablenotemark{a} } &
\multicolumn{2}{c}{\bf Energy Range (TeV)\tablenotemark{b}} \\
$\bf(deg)$ & \bf Sens. & {\bf Median}& {\bf 10\%-90\%} 
}
\startdata
0  & 6.5  & 40 & 9-150 \\    % -2.3 spectrum 
10 & 4.4  & 27 & 6-110 \\
20 & 3.6  & 22 & 5-82  \\
30 & 3.2  & 19 & 4-77  \\
40 & 3.1  & 19 & 4-77  \\
50 & 3.2  & 23 & 5-82  \\
60 & 3.5  & 26 & 6-100 \\
70 & 4.5  & 38 & 8-140 \\
\enddata
\tablenotetext{a}{ Flux sensitivity for a point source in units of $10^{-15}$TeV$^{-1}$s$^{-1}$cm$^{-2}$ quoted at 20 TeV for a $5\sigma$ detection.}
\tablenotetext{b}{ Energy below which the indicated percentage of weighted events are included in the analysis.}
\end{deluxetable}

% Source list table
\begin{deluxetable}{ccccccccc}
\tabletypesize{\tiny}  % can be \small \tiny \scriptsize and \footnotesize 
\tablewidth{0pt}
\tablecaption{Galactic Sources and Source Candidates\label{tab:galactic_sources}}
\tablehead{
\colhead{\bf Object} &
\colhead{\bf Location} &
\colhead{\bf Error\tablenotemark{a}} &
\multicolumn{3}{c}{\bf Significance($\sigma$)\tablenotemark{b}} &
\colhead{\bf Flux\tablenotemark{c} at 20 TeV} &
\colhead{\bf Extent}&
\colhead{\bf Counterparts} \\
    & $\bf (l,b)$ & {\bf Radius} & \bf pre- & \bf post-& $\bf 3^\circ\times3^\circ$ & $\bf \times10^{-15}$ &\bf Diameter& \bf (References)\\
    &             & {\bf (deg)}  & \bf trials & \bf trials     &           & $\bf TeV^{-1}cm^{-2}s^{-1}$ & \bf (deg) \\
}
\startdata
\bf Crab          & $ 184.5, -5.7$ & 0.11 & \bf 15.0 &\bf 14.3&11.5& $10.9\pm1.2$   &  -                      & Crab \\
\bf MGRO J2019+37 & $  75.0,0.2$   &      0.19 & \bf 10.4 &\bf 9.3 & 8.5& $8.7\pm1.4$ & $1.1^\circ\pm0.5^\circ$ \tablenotemark{d} & GEV J2020+3658, \\
                  &                &      &          &                 &    &             &                         & PWN G75.2+0.1, (1)\\
\bf MGRO J1908+06 & $  40.4, -1.0$ & 0.24 & \bf 8.3       &\bf 7.0 & 6.3& $8.8\pm2.4$ &$<2.6^\circ(90\%CL)$    & GEV J1907+0557,\\
                  &                &      &          &                 &    &             &                         & SNR G40.5-0.5\\
\bf MGRO J2031+41 & $  80.3,  1.1$ & 0.47 & \bf 6.6       &\bf 4.9 & 6.4&    $9.8\pm2.9$ & $3.0^\circ\pm0.9^\circ$ & GEV J2035+4214,\\
                  &                &      &          &                 &    &             &                         & TEV J2032+413 (2,3)\\
%\bf C1            & $  70.7,  1.0$ & \tablenotemark{e}& 4.6      & 2.0     & 3.9& $2.9\pm1.0$ & \tablenotemark{e}       & -  \\
\bf C1            & $  77.5, -3.9$ & 0.24 & 5.8      & 3.8 & 3.4& $3.1\pm0.6$ & $<2.0^\circ(90\%CL)$   & - \\
\bf C2            & $  76.1, -1.7$ & \tablenotemark{e}& 5.1      & 2.8     & 4.5& $3.4\pm0.8$ & \tablenotemark{e}       & - \\
\bf C3            & $ 195.7,  4.1$ & 0.40 & 5.1      & 2.8     & 5.9& $6.9\pm1.6$ & $2.8^\circ\pm0.8^\circ$ & Geminga\\
\bf C4            & $ 105.8,  2.0$ & 0.52 & 5.0      & 2.6     & 6.3& $4.0\pm1.3$ & $3.4^\circ\pm1.7^\circ$ & GEV J2227+6106\\
                  &                &      &          &                 &    &             &                         & SNR G106.6+2.9\\
\enddata
\tablenotetext{a}{The table lists statistical errors only. The systematic pointing error is $<0.3^\circ$.}
\tablenotetext{b}{The post trials significances account for the trials incurred in searching the 3800 square degree region.}
\tablenotetext{c}{The table lists statistical errors only. The systematic flux error is 30\%.}
\tablenotetext{d}{For this high significance detection, the extent was computed using only large events. See \citet{2007ApJ...658L..33A} for details.}
\tablenotetext{e}{Gaussian fit of excess failed for this candidate.}
\tablerefs{(1) Abdo et al. 2007, (2) Butt et al. 2003, (3) Anchordiqui et al. 2007.}
\end{deluxetable}

\end{document}